\documentclass[11pt]{llncs}

\usepackage{psfrag,graphicx}
\usepackage{amsmath,amssymb}

\newcommand{\be}{\begin{equation}}
\newcommand{\ee}{\end{equation}}
\newcommand{\bq}{\begin{eqnarray}}
\newcommand{\eq}{\end{eqnarray}}

\newcommand{\ket}[1]{\left | \, #1 \right\rangle}
\newcommand{\bra}[1]{\left \langle #1 \, \right |}

\begin{document}
\title{Universal quantum computation with a non-Abelian topological memory}
\author{James R. Wootton \and 
        Ville Lahtinen \and 
        Jiannis K. Pachos}
\institute{School of Physics and Astronomy, University of Leeds, \\ Woodhouse Lane, Leeds LS2 9JT, UK}

\maketitle       
\begin{abstract}

An explicit lattice realization of a non-Abelian topological memory is presented. The correspondence between logical and physical states is seen directly by use of the stabilizer formalism. The resilience of the encoded states against errors is studied and compared to that of other memories. A set of non-topological operations are proposed to manipulate the encoded states, resulting in universal quantum computation. This work provides insight into the non-local encoding non-Abelian anyons provide at the microscopical level, with an operational characterization of the memories they provide.

\end{abstract}

\section{\label{intro}Introduction}
Anyons are quasiparticles with topological, and therefore non-local, properties \cite{Wilczek,Pachos2}, that may be realized on two-dimensional systems \cite{Laughlin,Read,Freedman,Wen,Levin,Fendley,Wootton}. There has been a number anyon-based proposals for the storage and manipulation of quantum information. Many of these proposals deal with so-called Abelian anyons, encoding quantum information in quasiparticle occupancies \cite{Lloyd,me,qpl} or ground state degeneracies \cite{Dennis}. Others utilize cluster state quantum computation \cite{Raussendorf}. In all cases one obtains a topologically protected quantum memory, but this protection does not extend to the processing of the stored information.

Non-Abelian anyon models possess quasiparticles with more complex behaviour than their Abelian counterparts \cite{Kitaev2}. Specifically, local measurements on two such quasiparticles cannot determine how they will behave if brought together as composite object. This non-local degree of freedom, known as the fusion channel of the two anyons, is ideal to encode quantum information, protecting against local errors as long as the nature of the anyons is not affected. The energy gap associated with the anyons ensures that there is a threshold error rate before this may occur. Furthermore, processing the information contained in non-Abelian anyons is possible while remaining within the energy gap, and so has the same advantages as adiabatic quantum computation \cite{gold}.

Computational schemes with non-Abelian anyons are usually presented at an abstract level \cite{Bravyi,Georgiev}, while those using Abelian anyons are often more explicit \cite{Dennis,Raussendorf}. This means that, though non-Abelian schemes provide the most promising proposals for fault-tolerant quantum computation, it is Abelian schemes that are better understood in terms of their underlying systems. Here we propose a quantum memory using non-Abelian anyons of the $D(S_3)$ model, expressed explicitly in terms of the underlying spin lattice. This provides an opportunity to perform in-depth studies of the non-Abelian storage. Universal quantum computation is possible when the full $D(S_3)$ model is used \cite{mochon,Aguado}, but we restrict ourselves to a non-universal sub-model. This is because the memory is our primary concern, which can be more thoroughly studied when less anyon types are considered. It also gives us an opportunity to consider how to achieve universality by non-topological operations \cite{Raussendorf,bk,Bravyi,qpl}, and to see how they these work in terms of the underlying spins.

\subsection{The $D(S_3)$ anyon model}

Stabilizer codes, strictly defined, are based on lattices of two level spins and the corresponding Pauli group of operators \cite{Gottesman}. The quantum double models of anyons, proposed by Kitaev \cite{Kitaev2}, use a generalization of this concept. Spins of higher dimensions are employed, with operators based upon group structures. Abelian groups give rise to Abelian anyons, while non-Abelian groups lead to non-Abelian anyons. Here we consider the simplest non-Abelian model, $D(S_3)$, whose explicit lattice realization was outlined in \cite{Aguado}. This provides the tools with which we build our computational scheme. The relevant aspects of the model are summarized below.

The $D(S_3)$ anyon model is defined on an oriented two-dimensional square lattice. On each edge there resides a six-level spin spanned by the states $\ket{g}$, where $g$ is an element of $S_3$, the permutation group of three objects. We express every element in terms of generators $t$ and $c$, which satisfy $t^2 = c^3 = e$ and $tc = c^2 t$. $e$ denotes the trivial element. Using this notation the six elements are given by $S_3 = \{ e, c, c^2, t, tc, tc^2 \}$.

\begin{figure}[ht]
\begin{center}
{\includegraphics[scale=.5]{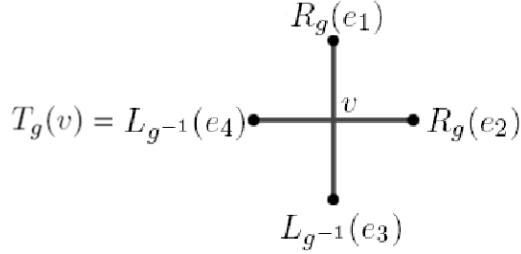} }
\caption{\label{fig1}A pictorial representation of the vertex operators $T_g (v)$.}
\end{center}
\end{figure}

Define a vertex operator acting on vertex $v$ by,
\be
T_g(v) = R_g (e_1) R_g (e_2) L_{g^{-1}} (e_3) L_{g^{-1}} (e_4), \quad [T_g(v), T_h(v')] = 0,
\ee
where the $e_i$ are the four edges connected to vertex $v$ (see Fig. \ref{fig1}). Here $R_g(e)$ and $L_g(e)$ denote the right and left multiplication, respectively, of the local spin state on edge $e$ by the element $g$. To be precise, they act as $R_g\ket{h} = \ket{hg}$ and $L_g\ket{h}=\ket{gh}$. For the purposes of our topological memory, we consider only the so-called charge anyons associated with the vertices of the lattice. There are two non-trivial charges, which we call $\Lambda$ and $\Phi$, and the trivial vacuum charge, $1$. When $\ket{\Psi}$ denotes a general state of the system, the presence of a charge of type $A$ at vertex $v$ is defined by $P_{A}\ket{\Psi}=\ket{\Psi}$, where the orthogonal projectors are given by,
\bq \label{p} \nonumber
P_{1}(v) &=& \frac{1}{6}[T_e (v) + T_c (v) + T_{c^2} (v) + T_t (v) + T_{tc} (v) + T_{tc^2} (v))], \\ \nonumber
P_{\Lambda}(v) &=& \frac{1}{6}[T_e (v) + T_c (v) + T_{c^2} (v) - T_t (v) - T_{tc} (v) - T_{tc^2} (v)], \\ \nonumber
P_{\Phi}(v) &=& \frac{1}{3}[2T_e (v) - T_c (v) - T_{c^2} (v)]. \nonumber
\eq
Projectors are also defined for the states of flux anyons on plaquettes, but we need not give them here.

The stabilizer space consists of states with no anyons, i.e. those for which $P_1(v)\ket{\textrm{gs}}=\ket{\textrm{gs}}$ for all $v$, and a similar condition for the fluxes on plaquettes. The syndrome measurement is defined as a measurement of anyon occupancies, and so corresponds to the above projectors. A Hamiltonian may be defined to maintain the stabilizer space. This assigns energy to the states of the anyons, and thus suppresses their spontaneous creation. This may be expressed,
\be
H = -\sum_v P_{1}(v) -\sum_p P_1 (p).
\ee

Charge anyons are created from the stabilizer space by acting with the following operators on single spins,
\bq
  W_{\Lambda}(e) & = & \ket{e}\bra{e} + \ket{c}\bra{c} + \ket{c^2}\bra{c^2} \nonumber \\
  \ & \ & \qquad - \ket{t}\bra{t} - \ket{tc}\bra{tc} - \ket{tc^2}\bra{tc^2}, \label{Wlambda} \\
  W_{\Phi}(e) & = & 2 \ket{e}\bra{e} - \ket{c}\bra{c} - \ket{c^2}\bra{c^2} \label{Wphi}.
\eq
These create charges on the two vertices connected by the edge $e$. A protocol to create and move charges several edges apart is given in \cite{Aguado}.

When charges of different type are brought to the same vertex, the possible outcomes are given by the fusion rules,
\be
\Lambda \times \Lambda = 1, \,\,\, \Lambda \times \Phi = \Phi, \,\,\, \Phi \times \Phi = 1 + \Lambda + \Phi.
\ee
The last implies that the $\Phi$ charges have three possible fusion channels; a pair may fuse to the trivial charge $1$, a $\Lambda$ or a $\Phi$. We may utilize the encoding of topological quantum computation \cite{Pachos2}, associating each possible outcome with a quantum state and hence using them to store quantum information. This information will be topologically protected due to the finite energy gap and the non-local encoding. However, the charges have trivial mutual statistics meaning that information processing by purely topological means is not possible. To achieve universal quantum computation, we propose non-topological operations to harness the power of the underlying spin lattice. As stated in  \cite{Bravyi}, abstract treatments of such quantum gates tend to be speculative. However, we have the means to study these gates explicitly in terms of spin operations.

\section{\label{lambda}The computation with $\Lambda$ charges alone}

Though we are using a stabilizer code, the encoding described above is not within the stabilizer space. This allows similar protection from errors, yet easier manipulation. The basic principles of our scheme for universal quantum computation are first presented using the $\Lambda$ charges alone. Topological protection is introduced later by encoding the $\Lambda$ charges within the fusion channels of $\Phi$'s, making the logical states indistinguishable by the stabilizer.

\begin{figure}[ht]
\begin{center}
{\includegraphics[scale=.75]{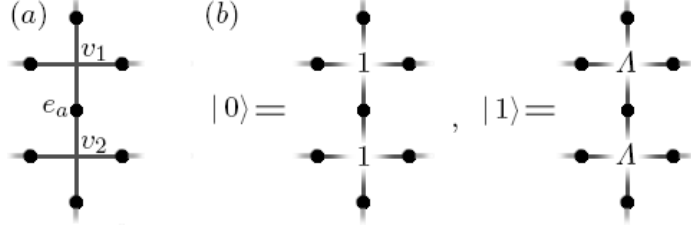} }
\caption{\label{fig2}(a) Two vertices use to store a logical qubit. (b) The logical states are stored by placing the trivial charge, $1$, or the charge $\Lambda$ at each vertex.}
\end{center}
\end{figure}

Consider two neighbouring vertices, $v_1$ and $v_2$, connected by the edge, $e_a$ (see Fig. \ref{fig2}(a)). The two vertices may be used to store a logical qubit $a$ by identifying trivial charge or a pair of $\Lambda$ charges at both $v_1$ and $v_2$ with the logical qubit states $\ket{0}_a$ and $\ket{1}_a$, respectively. Explicitly,
\bq \nonumber
\ket{0}_a &=& \ket{\rm gs}, \\
\ket{1}_a &=& W_{\Lambda}(e_a) \ket{\rm gs}.
\eq
These states are also expressed in Fig. \ref{fig2}(b).

Measurement in the $Z$ basis requires measurement of either vertex's occupancy, using the four-spin projectors of Eq.(\ref{p}). The logical $X$ is realized by $W_{\Lambda}(e_a)$, hence all operations diagonal in the $X$ basis act on the spin $e_a$ alone. The relation $\ket{\pm} \bra{\pm} = (I \pm X)/2$ may be used to write the $X$ basis projectors in terms of the lattice spins, 
\bq \nonumber
\ket{+}_a \bra{+} &=& \frac{I + W_{\Lambda}(e_a)}{2} =  \ket{e}_{e_a}\bra{e} + \ket{c}_{e_a}\bra{c} + \ket{c^2}_{e_a}\bra{c^2}, \\ 
\ket{-}_a \bra{-} &=& \frac{I - W_{\Lambda}(e_a)}{2} =  \ket{t}_{e_a}\bra{t} + \ket{tc}_{e_a}\bra{tc} + \ket{tc^2}_{e_a}\bra{tc^2}.
\eq
Measurement in the $X$ basis is therefore achieved by measuring the lattice spin in the above subspaces. Arbitrary phase gates in the $X$ basis may then be written,
\bq \nonumber
U_{\theta}(e_a) &=& \ket{+}_a \bra{+} + e^{i \theta} \ket{-}_a \bra{-} \\ \nonumber
&=& \Big(\ket{e}_{e_a} \bra{e} + \ket{c}_{e_a}\bra{c} + \ket{c^2}_{e_a} \bra{c^2} \Big) \\ 
&+& e^{i \theta} \Big(\ket{t}_{e_a} \bra{t} + \ket{tc}_{e_a}\bra{tc} + \ket{tc^2}_{e_a}\bra{tc^2} \Big).
\eq
These may be easily performed with single spin rotations.

Entanglement with another logical qubit, $b$, stored on another pair of vertices with shared spin $e_b$, may be achieved by the phase-controlled-NOT gate. This is diagonal in the $X$ basis of both qubits, and acts only on $e_a$ and $e_b$. It may be expressed as follows,
\be
K_{a,b} = \ket{+}_a \bra{+} \otimes I_b + \ket{-}_a \bra{-} \otimes X_b = I + W_{\Lambda}(e_a) + W_{\Lambda}(e_b) - W_{\Lambda}(e_b) W_{\Lambda}(e_b).
\ee
These operations form a universal gate set for quantum computation. For example, a Hadamard may be implemented on an arbitrary state $\ket\psi = \alpha \ket{+}_a + \beta \ket{-}_a$ of qubit $a$ as follows. Firstly, qubit $b$ is prepared in the state $\ket{0}_b$, and then entangled to $a$ using $K_{a,b}$. The resulting state is,
\bq \nonumber
\alpha \ket{+0}_{a,b} + \beta \ket{-1}_{a,b}  &=& \frac{1}{\sqrt{2}} \Big( \alpha \ket{00}_{a,b}  + \alpha \ket{10}_{a,b}  + \beta \ket{01}_{a,b}  - \beta \ket{11}_{a,b} \Big) \\
&=& \frac{1}{\sqrt{2}} \Big( \ket{0}_a (H\ket{\psi}_b) + \ket{1}_a (ZH\ket{\psi}_b) \Big).
\eq
Measuring qubit $a$ in the $Z$ basis then yields the state $H \ket{\psi}$ on qubit $b$, followed by a $Z$ if the outcome of the measurement is $\ket{1}_a$. In the latter case the process may be repeated until the error is corrected and a Hadamard alone is implemented. With this Hadamard and the arbitrary phase gates in the $X$ basis, arbitrary single qubit unitaries may be performed. With the entangling gate, this leads to universal quantum computation \cite{ike+mike}.

\section{\label{phi}Fault-tolerance using non-Abelian charges}

We will now extend the encoding by using $\Phi$ charges to hide the $\Lambda$'s. We first consider the most straightforward way of doing this, and then explore an alternative method.

\begin{figure}[ht]
\begin{center}
{\includegraphics[scale=1]{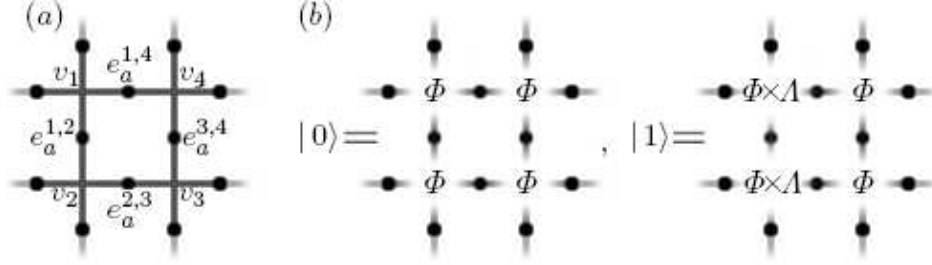} }
\caption{\label{fig3}(a) Four vertices use to store a logical qubit. These are labelled from $v_1$ to $v_4$, starting from the top left and proceeding anticlockwise. The spin along the side connecting the vertices $v_i$ and $v_j$ is denoted $e^{i,j}_a$. (b) Both logical states use a $\Phi$ charge at each vertex. The only difference is that two of these are fused with a $\Lambda$ charge in the $\ket{1}$ state. There is no local way to detect this, especially when the charges are separated.}
\end{center}
\end{figure}

Let us consider four neighbouring vertices, as shown in Fig. \ref{fig3}(a). Pairs of $\Phi$ charges carrying the trivial fusion channel may be created from the ground state with $W_{\Phi}$ (\ref{Wphi}). 
Applying this to spins $e^{1,4}_a$ and $e^{2,3}_a$ creates a pair carrying the trivial fusion channel on $v_1$ and $v_4$, and another on $v_2$ and $v_3$. This state is identified with the logical qubit state $\ket{0}_a$. By applying $W_{\Lambda}(e^{1,2}_a)$, a $\Lambda$ charge is fused with a $\Phi$ from each pair again resulting in two $\Phi$ pairs except that they now belong to the $\Lambda$ fusion channel. This state is identified with the logical qubit state $\ket{1}_a$. Explicitly,
\bq \label{log} \nonumber
\ket{0}_a &=& W_{\Phi}(e^{1,4}_a) W_{\Phi}(e^{2,3}_a) \ket{\rm gs}, \\
\ket{1}_a &=& W_{\Phi}(e^{1,4}_a) W_{\Phi}(e^{2,3}_a) W_{\Lambda}(e^{1,2}_a) \ket{\rm gs}.
\eq
These states are also expressed in Fig. \ref{fig3}(b). Further logical qubits may be stored on other sets of four $\Phi$ charges. The syndrome measurements will see only the $\Phi$ charges and not the $\Lambda$ charges they contain, making the logical states indistinguishable by local measurements alone, and degenerate under the Hamiltonian.

We observe that $W_{\Lambda} W_{\Phi} = W_{\Phi}$, implying,
\be
W_{\Lambda}(e^{1,2}_a) W_{\Lambda}(e^{1,4}_a) W_{\Phi}(e^{1,4}_a) = W_{\Lambda}(e^{1,2}_a) W_{\Phi}(e^{1,4}_a).
\ee
Here the left-hand side creates a $\Phi$ pair on $v_1$ and $v_4$ and fuses a $\Lambda$ with the $\Phi$ on $v_1$. The right-hand side does the same except that the $\Lambda$ is fused with the $\Phi$ on $v_4$. The equality between these shows that the resultant state does not depend upon which $\Phi$ the $\Lambda$ was fused with, and holds even when they are well-separated, showing that the encoding of information in this way is indistinguishable by local operators alone. 

Rather than keeping the $\Phi$ charges on neighbouring vertices, it is possible to move them apart. The single spins $e^{i,j}_a$ are then replaced by chains $C^{i,j}_a$ of $l$ spins, where $l$ is the new separation between the anyons. The logical states will be similar in form to those of Eq.(\ref{log}) except that operations acting on spins $e^{i,j}_a$ will instead act on the chains $C^{i,j}_a$. The operations $W_{\Phi}[C^{i,j}_a]$ take the form,
\be
W_{\Phi}[C^{i,j}_a] = \sum_{g_n\times...\times g_1=c^k} (\omega^k + \omega^{-k}) \ket{g_1,...,g_n}\bra{g_1,...,g_n},
\ee
where $g_1,...,g_n$ are the states of the spins within the chain $C^{i,j}_a$ and $\omega=e^{i 2 \pi /3}$. The operations $W_{\Lambda}[C^{i,j}_a]$ are simply the product of $W_{\Lambda}$ on each spin along a the chain.  Just as in the previous section, this operation provides the logical $X$. Hence all $X$ basis operations determined there still apply unchanged, except that they must now act on $O(l)$ spins to be realized. Measurement in the $Z$ basis now requires the fusion of one or other of the $\Phi$ pairs and measurement of the result, the trivial charge implying $\ket{0}_a$ and $\Lambda$ implying $\ket{1}_a$. These operations achieve universal quantum computation in the same way as before.

Errors in the encoding come from fusion with stray charges or braiding with stray fluxes. Both are suppressed by the stabilizer code, since regular measurements of the syndrome can detect these anyons and allow for their annihilation. They are also suppressed by the Hamiltonian, since the creation of the stray anyons costs energy. To see how well the errors are suppressed, we will now consider them individually. Errors in the $X$ basis are caused by the creation of stray $\Lambda$ charges and their fusion with a $\Phi$ from each pair. This requires a string of errors to occur on the $l$ spins between the $\Phi$'s, a process whose probability is suppressed by $O(e^{-l})$  \cite{Kitaev2,Dennis}. Since the size of the logical operations only increases linearly with $n$, this is an efficient suppression of errors. Errors in the $Z$ basis come from fusion with stray $\Phi$'s, which can disrupt those used to encode and thus leave the logical information exposed to the stabilizer, and lose its degeneracy under the Hamiltonian. $Z$ basis errors can also come from braiding with stray fluxes. Additional protection can be given to this basis by using a repetition code, in which two sets of $n$ $\Phi$ charges are used to encode each qubit, rather than just two pairs \cite{qpl}. The probability of errors will then be suppressed by $O(e^{-n})$.

It is possible to move the $\Phi$ charges using either multi-spin operations \cite{Aguado} or local potentials \cite{me}. This gives the scheme a useful flexibility, since the charges may be moved apart to harness improved protection against errors and moved close so that logical operations may be performed on less spins. 

\subsection{Relation to other topological memories}

The $\Lambda$ occupation of a vertex can be determined by measuring the observable $T_t (v)$, with the presence of the charge signalled by an outcome of $-1$. This is true even when a $\Phi$ is present on the vertex, since the measurement can even detect those $\Lambda$'s fused with $\Phi$'s. Consequently, making such a measurement on two $\Phi$ charges allows us to determine the number of $\Lambda$'s within the $\Phi$ pair. As one might expect, an even number will be found within any $\Phi$ pair that will fuse to vacuum, since the $\Lambda$'s will annihilate upon fusion. An odd number will be found within any $\Phi$ pair that will fuse to a $\Lambda$. The LOCC protocol of measuring $T_t (v)$ on each $\Phi$ and collecting the results is therefore sufficient to distinguish the logical states of Eq.(\ref{log}). Note that since these measurements only act on the spins directly around each $\Phi$, increasing their separation will not affect the complexity of the protocol.

Consider a modification to the syndrome measurement, in which the projectors $P_{\Lambda}(v)$ are replaced by $P'_{\Lambda}(v)=[T_e (v) + T_t (v)]/2$, and can therefore detect the $\Lambda$'s within $\Phi$'s. Since the syndrome measures each vertex and collects the results, it is able to count the number of $\Lambda$'s within each $\Phi$ pair, and thus distinguish the logical states. This shows that the encoding is equivalent to those in which Abelian anyons are stored in holes \cite{Raussendorf,qpl}, since using the standard syndrome is equivalent to using the modified syndrome with the $P'_{\Lambda}(v)$ projections suppressed on any vertex holding a $\Phi$. A exciting implication of this is that Abelian models may be used for non-Abelian-like encodings, using the principles of non-Abelian anyons to enhance the power of their memories \cite{qna,thesis}.

To see how a stronger encoding may be constructed, let us consider the single spin operation,
\be \label{w'}
W'_{\Phi} = \ket{c}\bra{c} - \ket{c^2}\bra{c^2}.
\ee
Like $W_{\Phi}$, this creates $\Phi$ charges on the vertices either side of the spin. However, measurements of $T_t (v)$ will give different results. An odd number of $\Lambda$'s will be found within a pair of $\Phi$ charges that fuse to vacuum, and an even number found within those that fuse to a $\Lambda$. This is opposite to what one would expect. The relative minus sign, coupled with the non-Abelian group multiplication underlying all operations on the spins, causes the $T_t (v)$'s to detect one more $\Lambda$ within a pair than is actually present. Using this property, the logical states may be made indistinguishable to the $T_t (v)$ measurements, and any LOCC protocol, by using differently defined $\Lambda$ pairs for the logical states. Explicitly,
\bq \label{log2} \nonumber
\ket{0}_a &=& W_{\Phi}(e^{1,4}_a) W_{\Phi}(e^{2,3}_a) \ket{\rm gs}, \\
\ket{1}_a &=& W'_{\Phi}(e^{1,4}_a) W'_{\Phi}(e^{2,3}_a) W_{\Lambda}(e^{1,2}_a) \ket{\rm gs}.
\eq
With this encoding an even number of $\Lambda$'s is found within any pair, regardless of their fusion channel. They are then distinguishable only with non-local operations, such as the fusion of $\Phi$'s. This is the true non-Abelian encoding, whose protection goes above and beyond that of Abelian encoding with holes. 

Note that the huge operational difference between this encoding and that of Eq. (\ref{log}) comes directly from the non-Abelian group multiplication underlying the model. It is only because of this that the relative minus sign in Eq. (\ref{w'}) has such an effect. Abelian group multiplication cannot provide tricks to fool the $T_t (v)$ observables in such a way.

The stronger encoding increases the complexity of the logical $X$ operation. The fusion of a $\Phi$'s with a $\Lambda$'s is no longer enough. The unitary operation,
\be
U(v) = \frac{1}{3} T_e (v) - \frac{2}{3}[\omega \, T_c (v) + \omega^2 \, T_{c^2} (v)],
\ee
must be applied to any vertex on which a fusion takes place to rotate from $W_{\Phi}$ type $\Phi$ pairs to $W'_{\Phi}$ type, or vice-versa. Rather than single spin operations, logical operations on neighbouring $\Phi$'s must then act on seven spins. For non-neighbouring $\Phi$'s, operations must also act on six more spins than the previous requirement. Though the size of logical operations still scales with $O(l)$, and so still gives efficient suppression of errors, it is not as accessible to actual experimental realization. 

\section{\label{end}Conclusions and further work}

We have proposed a novel scheme for fault-tolerant quantum computation, utilizing a non-Abelian topological memory. As a result of this work, we have an explicit form for the logical states stored non-locally in terms of the physical states of the underlying lattice model, an understanding of what kinds of memories are possible and their relations to other topological memories. Specifically, we have found two means to encode qubits in the fusion channels of the model's anyons. Though both fault-tolerant and indistinguishable to local operations, these encodings have a crucial difference. One has states distinguishable to LOCC protocols, and is equivalent to encodings using Abelian anyons. The other has states distinguishable only to non-local operations. Hence, by showing exactly how these encodings differ, we have demonstrated the true difference between Abelian and non-Abelian anyons from a quantum information perspective.

Furthermore, we harness these states to give the non-topological operations required for universality while remaining below the energy gap. Our work allows the application of realistic error models and studies of how anyonic systems respond to practical experimental conditions \cite{topent}. There exist proposals on how to realize this and other lattice models in the laboratory \cite{Aguado,Doucot,Pachos,Yang,Zoller}. This exercise is a step towards physical realizations of simple non-Abelian systems to demonstrate aspects of quantum computation.

We also note that the use of single spin measurements on highly entangled states bears a similarity to measurement based quantum computation \cite{Briegel}. It would be beneficial to unify these formalisms.

\section*{Acknowledgements}

We would like to thank Gavin Brennen for inspiring conversations. This work was supported by the EU grants SCALA and EMALI, the EPSRC, the Finnish Academy of Science and the Royal Society.

\end{document}